\documentclass[aps,prb,reprint,amssymb,groupedaddress]{revtex4-1}
\usepackage{graphicx}% Include figure files
\usepackage{epstopdf}%fix the fact that pdflatex doesnt like *.eps files
\usepackage{dcolumn}% Align table columns on decimal point
\usepackage{bm}% bold math
\usepackage{color}
\usepackage{amsmath}
\begin{document}
\title[Crystal field states in a Quantum Spin Ice]{Crystal field states of Pr$^{3+}$ in the candidate quantum spin ice Pr$_{2}$Sn$_{2}$O$_{7}$}

\author{A. J. Princep}
\email[]{a.princep@physics.ox.ac.uk}

\author{D. Prabhakaran}

\affiliation{Department of Physics, University of Oxford, Clarendon Laboratory,
Parks Road, Oxford, OX1 3PU, United Kingdom}

\author{D. T. Adroja}

\affiliation{ISIS Facility, Rutherford Appleton Laboratory, STFC, Chilton, Didcot,
Oxon, OX11 0QX, United Kingdom}

\author{A. T. Boothroyd}
\email[]{a.boothroyd@physics.ox.ac.uk}

\affiliation{Department of Physics, University of Oxford, Clarendon Laboratory,
Parks Road, Oxford, OX1 3PU, United Kingdom}

% repeat the \author .. \affiliation  etc. as needed
% \email, \thanks, \homepage, \altaffiliation all apply to the current
% author. Explanatory text should go in the []'s, actual e-mail
% address or url should go in the {}'s for \email and \homepage.
% Please use the appropriate macro foreach each type of information

% \affiliation command applies to all authors since the last
% \affiliation command. The \affiliation command should follow the
% other information
% \affiliation can be followed by \email, \homepage, \thanks as well.
\author{}
%\email[]{Your e-mail address}
%\homepage[]{Your web page}
%\thanks{}
%\altaffiliation{}
\affiliation{}

\date{\today}

\begin{abstract}

Neutron time-of-flight spectroscopy has been employed to study the crystal-field splitting of Pr$^{3+}$ in the pyrochlore stannate Pr$_{2}$Sn$_{2}$O$_{7}$. The crystal field has been parameterized from a profile fit to the observed neutron spectrum.
The single-ion ground state is a well isolated non-Kramers doublet of $\Gamma_3^+$ symmetry with a large Ising-like anisotropy, $\chi_{zz} /\chi_\perp \approx 60$ at 10 K, but with a significant admixture of terms $| M_J \neq \pm J \rangle$ which can give rise to quantum zero-point fluctuations.
This magnetic state satisfies the requirements for quantum spin ice behavior.

\end{abstract}

% insert suggested PACS numbers in braces on next line
% insert suggested keywords - APS authors don't need to do this
%\keywords{}

%\maketitle must follow title, authors, abstract, \pacs, and \keywords
\maketitle

\section{Introduction}

Magnetic moments in the pyrochlores A$_2$B$_2$O$_7$ ($A = $ rare earth, $B = $ Ti, Zr, Ir, Sn, ...) are highly geometrically frustrated, and known to show complex magnetic ground states ranging from spin liquids to spin glasses and spin ices \cite{pyrochloresreview, spinliquidsmonopoles}. This diversity of phenomena arises out of the interplay between crystal field, magnetic exchange and dipolar interactions. When the dipolar interactions dominate over exchange interactions in systems with strong Ising-like anisotropy in the $\langle 111\rangle$ directions, spin-ice ground states are realized, for example in Ho$_2$Ti$_2$O$_7$ and Dy$_2$Ti$_2$O$_7$. Spin ice is a very special magnetically frustrated ground state in which the spins at the corners of each tetrahedron on the pyrochlore lattice freeze into a two-in-two-out configuration, analogous to the proton correlations in water ice.
 Interest in spin ice materials has burgeoned thanks to the prediction and subsequent experimental detection of magnetic monopole-like quasiparticles, which are the fundamental excitations of the spin-ice ground state \cite{monopoleprediction, monopolediffusescattering, HoTiOmonopoles, monopolerelaxation}.

 In recent years there has been growing interest in the possibility of a state known as dynamic, or quantum, spin ice \cite{photonprediction}. Whereas the dynamics of classical spin ice slow down and eventually freeze as $T \rightarrow 0$, a quantum spin ice exhibits significant residual transverse spin fluctuations even at the lowest temperatures. If the nature and strength of the exchange interactions between spins is favourable, the fluctuations can become correlated allowing quantum mechanical tunneling within the ice rules manifold of states. It has been predicted that this particular spin liquid state could realise a fully dynamical, lattice analogue of quantum electromagnetism with linearly dispersing magnetic excitations exactly analogous to photons \cite{quantumspiniceexcitations, seeingthelight} in addition to other exotic excitations \cite{spinicestrings, YTOhiggs}.

  Most discussion of possible real-world candidates for quantum spin ice has concerned the titanates Tb$_2$Ti$_2$O$_7$ and Yb$_2$Ti$_2$O$_7$, both of which exhibit spin liquid features according to several different experimental probes  \cite{ytterbiumspinice, terbiumspinice}. At the same time, arguably the most promising candidates for observation of strong quantum effects are pyrochlores containing Pr$^{3+}$, because the large ionic radius and small moment of Pr$^{3+}$ enhances the exchange coupling and reduces the nearest-neighbour dipolar interaction relative to the heavy rare earths \cite{nonkramerspseudospin, genericquantumice, onodaPRB}. Another distinction of Pr systems is that quadrupolar interactions are expected to be important, and these could lead to new types of complex ground states, including states with non-trivial chiral correlations  (e.g. in  Pr$_2$Ir$_2$O$_7$, Ref.  \onlinecite{PrIrOspinliquid} ).

  Evidence has been found for a dynamic spin ice state at low temperatures in Pr$_{2}$Sn$_{2}$O$_{7}$, in which the nearest-neighbour dipolar interaction strength $D \approx 0.13$\,K is considerably weaker than the estimated exchange energy $J \approx 0.9$\,K \cite{MatsuhiraJMMM, ZhouPSO}. The zero-point entropy of Pr$_{2}$Sn$_{2}$O$_{7}$ is about $25\%$ higher than that of the Ho/Dy-based spin-ices, which indicates that the spins are much more dynamic than in a classical dipolar spin ice. The dynamic nature of the spins was confirmed by observations of the quasielastic width in high resolution neutron spectra, which revealed that significant relaxation persists down to temperatures as low as 0.2\,K.

 The feature that allows Pr$^{3+}$ pyrochlores to exhibit zero-point fluctuations is the presence of terms with $|M_J \neq \pm J \rangle$ in the ground state wave function imposed by the crystal field interaction \cite{nonkramerspseudospin, onodaPRB}. Susceptibility measurements suggest that the crystal field ground state of Pr$^{3+}$ in Pr$_{2}$Sn$_{2}$O$_{7}$ is a non-Kramers doublet with strong Ising-like single-ion anisotropy, and the measured low energy neutron spectra indicate that the first excited state is about 18 meV above the ground state. However, up to now there has been no direct determination of the crystal field interaction in Pr$_{2}$Sn$_{2}$O$_{7}$ or indeed in any other Pr-containing pyrochlore.

 In this work we used time-of-flight neutron inelastic scattering to measure the spectrum of single-ion excitations of Pr$^{3+}$ in Pr$_{2}$Sn$_{2}$O$_{7}$ up to 500\,meV. We use a detailed model of the Pr single-ion states, including intermediate coupling and $J$-mixing, to determine the single-ion Hamiltonian and hence to calculate the magnetic properties. The analysis shows that, as expected, the ground state is a non-Kramers doublet with a strong Ising-like anisotropy, and that it contains a significant admixture of terms  with $|M_J \neq \pm J \rangle$ . The results reinforce the view that Pr$_2$Sn$_2$O$_7$ is a strong candidate for quantum spin ice.

\section{Experimental Details}

 Polycrystalline Pr$_{2}$Sn$_{2}$O$_{7}$ and Y$_{2}$Sn$_{2}$O$_{7}$ samples were prepared by standard solid-state synthesis techniques as described in \cite{spinicepreparationmethod}. The Y$_{2}$Sn$_{2}$O$_{7}$ sample was used as a non-magnetic reference. Both samples were characterised via laboratory x-ray diffraction at room temperature and found to be single phase with the Pr$_{2}$Sn$_{2}$O$_{7}$ lattice parameter refined to be 10.60708(3) \AA. The Pr sample was additionally characterised by SQUID magnetometry, indicating an effective moment of 2.6 $\mu_B$ and a curie-weiss temperature of 0.3K.

For the neutron scattering measurements, approximately 15g of powder samples were enclosed in a thin wrapper of aluminium foil in a cylindrical geometry (diameter approximately 15mm), then mounted into a closed-cycle refrigerator on the neutron time-of-flight spectrometer MERLIN at the ISIS spallation source \cite{MERLINpaper}. Spectra were recorded for both samples with neutron incident energies $E_i=100$, $170$ and $550$\,meV and at temperatures $T = 5$, 100, 200 and 300 K. The raw data were normalized and corrected for detector efficiency and time-independent background following standard procedures.   Vanadium spectra recorded at the same set of incident energies were used to determine the energy resolution and to convert the intensities into units of cross section, mb sr$^{-1}$ meV$^{-1}$ f.u.$^{-1}$, where f.u. stands for formula unit of Pr$_{2}$Sn$_{2}$O$_{7}$. We estimated the non-magnetic contribution to the low-angle Pr$_{2}$Sn$_{2}$O$_{7}$ spectra by scaling the high-angle Pr spectrum by the ratio of the low- to high-angle Y spectra \cite{neutroncorrections}.

\section{Results and Analysis}

\begin{figure}
\includegraphics[clip,width=0.95\columnwidth]{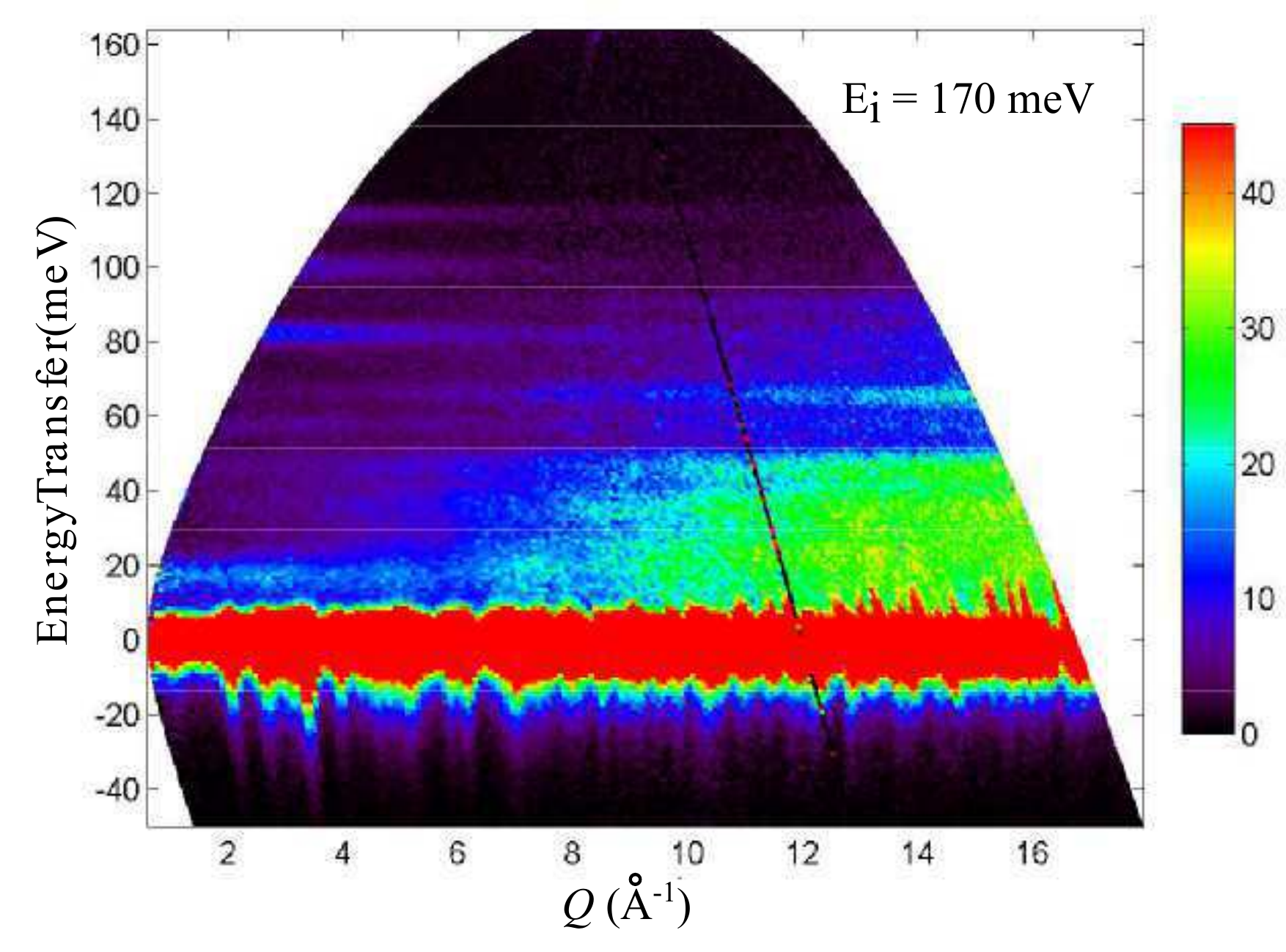}
\caption{\label{TOFdata} (Color online) Inelastic spectrum of polycrystalline Pr$_{2}$Sn$_{2}$O$_{7}$ as a function of scattering vector $Q$. The spectrum was recorded on the time-of-flight neutron spectrometer MERLIN with an incident energy $E_i = 170$\,meV. The CF transitions are the bands of scattering observed at low $Q$ whose intensities decrease with $Q$. The strong scattering at large $Q$ is from phonons. The intensity is in units of mb sr$^{-1}$ meV$^{-1}$ f.u.$^{-1}$.}
\end{figure}

\begin{figure}
\includegraphics[width=0.5\textwidth]{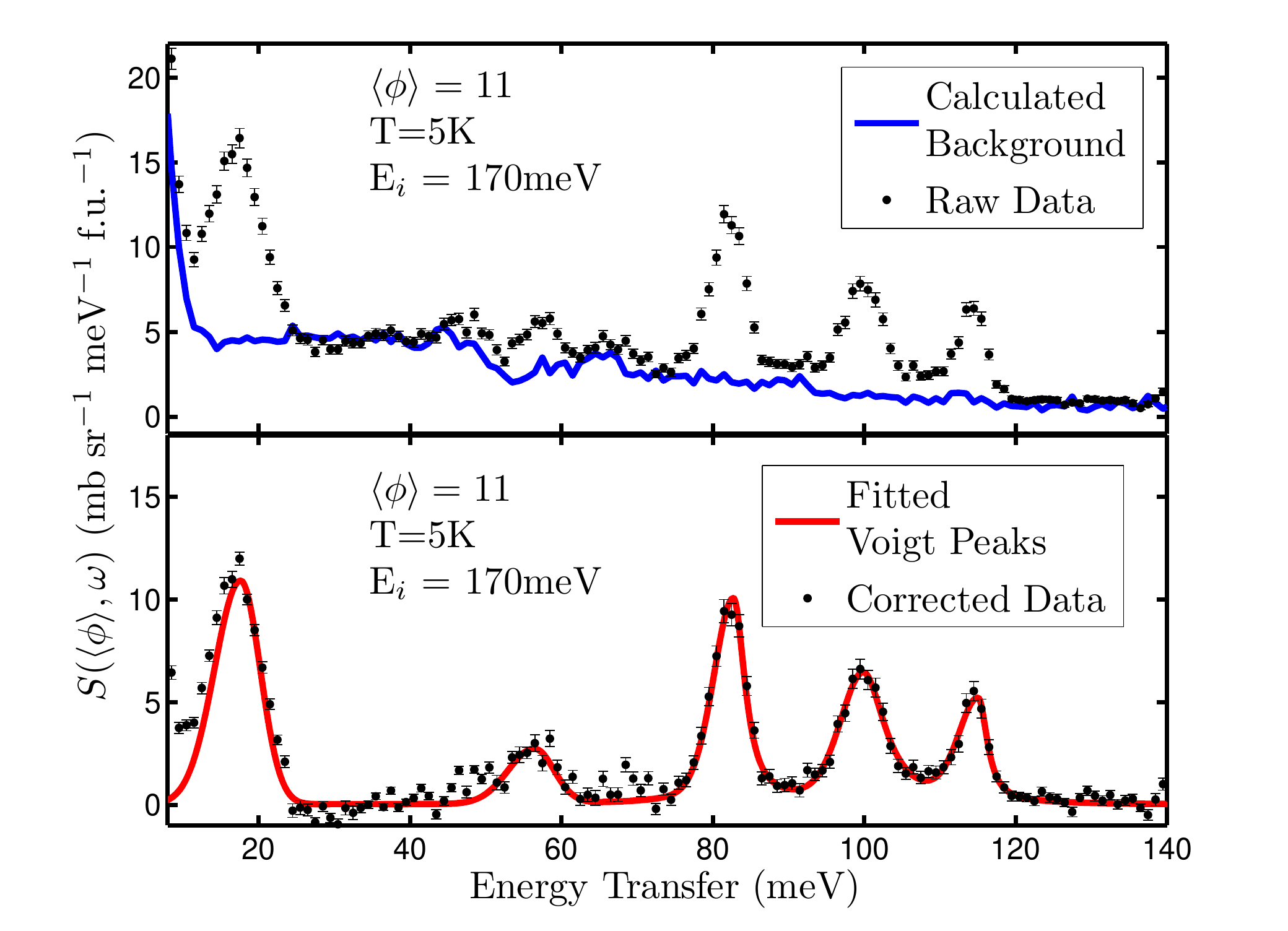}
\caption{\label{dataandfit1} Neutron scattering intensity $S(\langle \phi \rangle, \omega)$ as a function of energy transfer / momentum transfer, measured with an incident energy $E_i=170$meV. The upper panel shows the raw spectrum together with the non-magnetic background estimated by the method outlined in the text. The lower panel shows the background-corrected spectra fitted with a series of asymmetric pseudo-voigt peaks. }
\end{figure}

 Figure \ref{TOFdata} shows the raw time-of-flight spectrum of Pr$_{2}$Sn$_{2}$O$_{7}$ measured with an incident energy of 170\,meV, as a function of the magnitude of the scattering vector $Q$. The strong signal centered on zero energy is from elastic scattering, and the general increase in intensity with $Q$ is caused by scattering from phonons. Crystal-field transitions cause the weak dispersionless bands of scattering seen most clearly at small $Q$. Figure \ref{dataandfit1}  shows energy spectra integrated over a small range of scattering angle $\phi$. The average scattering angle weighted by solid angle is denoted by $\langle \varphi \rangle$. In the lower panels of  Fig.~\ref{dataandfit1} the solid line shows asymmetric pseudo-Voigt functions that were fitted to the background-corrected data in order to obtain the transition energies and integrated intensities. The results of these fits are listed in Table \ref{table1}, where the integrated intensities are recorded relative to the largest peak, i.e. the one at 17.8\,meV, and are extrapolated to Q=0 by correcting for the dipole magnetic form factor.

\begin{figure}
\includegraphics[width=0.45\textwidth]{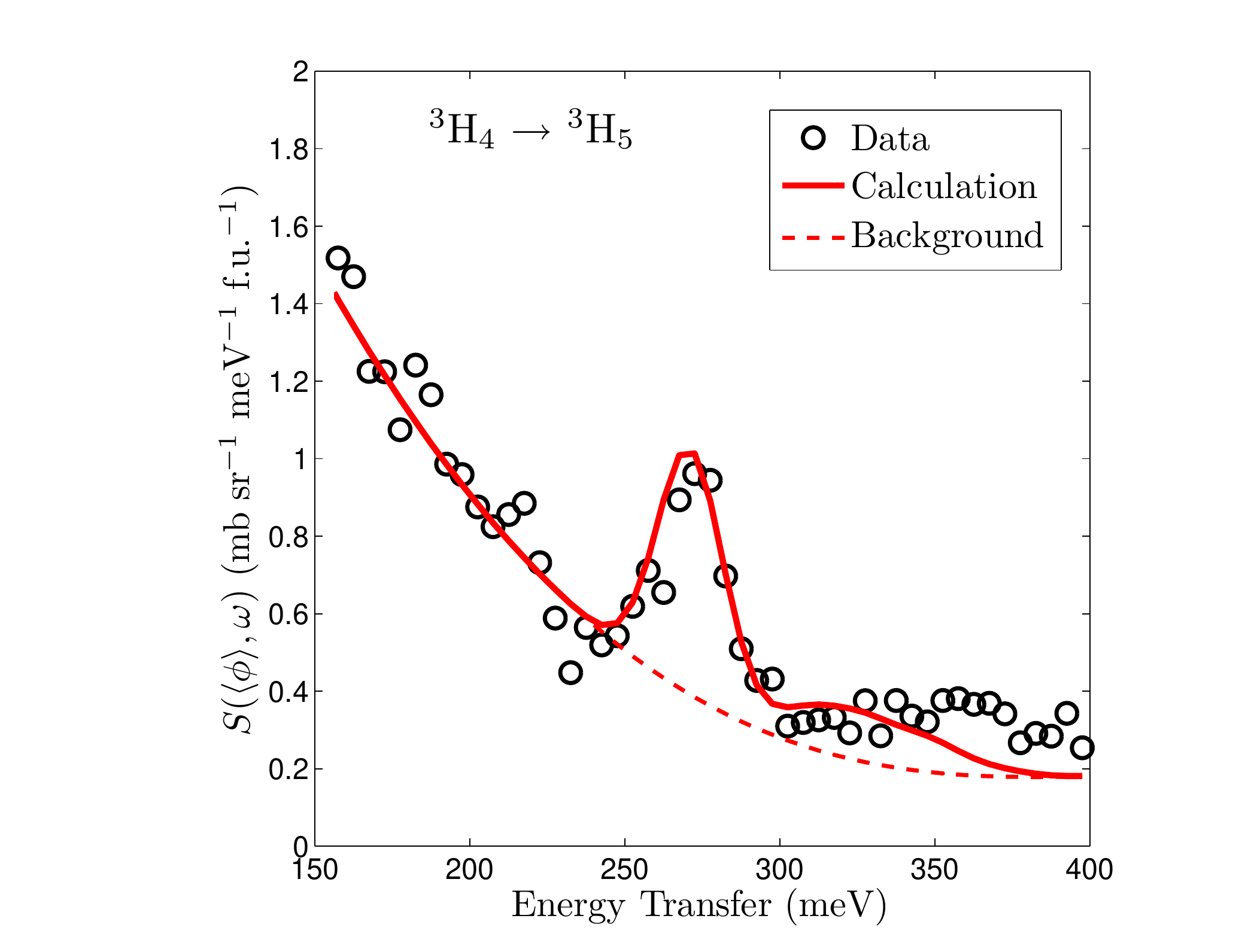}
\caption{\label{intermultiplet} Neutron scattering spectrum of $^3H_4$ $\rightarrow$ $^3H_5$ intermultiplet transitions. The data were recorded at a temperature 5\,K, and plotted for a mean solid angle $\langle \phi \rangle = 7.56^\circ$.The broken line is an estimate of the unstructured background, and the solid line is calculated from the crystal field model without scaling. The excess signal centered on an energy transfer of approximately 360\,meV is attributed to a hydrogen vibrational mode.}
\end{figure}

 The spectrum in Fig.~\ref{intermultiplet} covers the energy range where transitions to the crystal-field split $^3H_5$ multiplet are expected. The spectrum contains a prominent peak centred on 270\,meV with a high energy shoulder extending to about 400\,meV. The shoulder feature is observed to extend to high $Q$ (not shown), and so is attributed to a vibrational mode, most likely from a small amount of hydrogen-containing contaminant in the powder sample formed through exposure to air.

 The crystal field acting on the Pr ions in Pr$_{2}$Sn$_{2}$O$_{7}$ has point symmetry $\overline{3}m$ ($D_{3d}$), with the local $\overline{3}$ axes parallel to the $\langle 111 \rangle$-type directions of the crystal lattice. We assume that the spectrum can be described by transitions between localized states of the pure $4f^2$ configuration of Pr$^{3+}$, and employ intermediate coupling basis states. In the absence of the crystal field interaction, the intermediate coupling basis states are dominated by the Hund's rule ground state $^3H_4$ $(L=5, S=1, J=4)$, with a small admixture of $^3F_4$ and $^1G_4$ (Ref. \onlinecite{prmultipletneutrons}).  The crystal field then splits the $J=4$ intermediate coupling ground multiplet into three doublets and three singlets, with symmetry decomposition $3\Gamma^+_3 + 2\Gamma^+_1 + \Gamma^+_2$ in terms of the irreducible representations of $\overline{3}m$ ($D_{3d}$) \cite{Kosterpointsymmetry}.  With the quantization axes along the $\overline{3}$ axis the CF Hamiltonian takes the form
\begin{eqnarray}
\mathcal{H}_{\rm CF} = & B^2_0C^2_0 + B^4_0C^4_0 + B^4_3(C^4_{-3}-C^4_3) + B^6_0C^6_0 \nonumber \\
&+ B^6_3(C^6_{-3}-C^6_3)+ B^6_6(C^6_{-6}+C^6_6),
\label{eq:Hamiltonian}
\end{eqnarray}
where $B_q^k$ denote the crystal field parameters (CFPs) and $C_q^k$ are the components of the tensor operator $C^k$ (Ref. \onlinecite{Wybournebook}). The strength of the CF interaction in Pr$_{2}$Sn$_{2}$O$_{7}$ is comparable to the spin--orbit interaction in Pr, so we expect significant $J$-mixing in the crystal-field wavefunctions. We therefore diagonalised $\mathcal{H}_{\rm CF}$ within the complete set of 91 intermediate coupling basis states of the $f^2$ configuration of Pr$^{3+}$ using the program \emph{Spectre}\cite{spectre}.

The measured neutron scattering cross section is proportional to the response function of the system, which for single-ion magnetic transitions is given by\cite{neutronbook}
\begin{eqnarray}
S({\bf Q},\omega) & = & \left(\frac{\gamma r_0}{2}\right)^{\hspace{-2pt}2}{\rm e}^{-2W}\sum_{i}p_{i}\sum_{j} \nonumber \\
& & |\langle\Gamma_j|\,{\bf M}_{\perp}({\bf Q})\,|\Gamma_i\rangle|^2\delta(E_{j} - E_{i}-\hbar\omega),
\label{eq:S(Q,w)_CF_transitions}
\end{eqnarray}
where $\bf Q$ is the scattering vector, $\hbar \omega$ is the energy transferred from the neutron to the sample, and $(\gamma r_0/2)^2 = 72.7$\,mb. The first summation is over the initial states $\Gamma_i$ with thermal population $p_i$, and the second summation is over the final states $\Gamma_j$. We assume the Debye--Waller factor ${\rm e}^{-2W}$ to be unity at the low temperatures of the measurements. We calculated the powder-averaged intensities of transitions within the lowest $J$ multiplet using the dipole approximation, in which the magnetic scattering operator may be written
\begin{equation}
 {\bf M}_{\perp}({\bf Q}) = -f(Q)g_J{\bf J}_{\perp},
 \label{M_perp_dipole_approx}
\end{equation}
where $f(Q)$ is the dipole form factor and ${\bf J}_{\perp}$ is the component of $\bf J$ perpendicular to $\bf Q$. For the intermultiplet transition intensities we used the full multipole expression for the scattering operator \cite{neutronbook}.

 We determined the crystal field model for Pr$_{2}$Sn$_{2}$O$_{7}$ by varying the CFPs in $\mathcal{H}_{\rm CF}$, eqn~(\ref{eq:Hamiltonian}), until best agreement was obtained between the measured and calculated energies and relative intensities for transitions within the $J=4$ ground state multiplet (up to $\sim 120$\,meV). As a starting point, we took the crystal field model determined for Ho$_{2}$Ti$_{2}$O$_{7}$ by S. Rosenkranz et. al. ( Ref. \onlinecite{HoTOCEF}) and scaled the CFPs for Ho$^{3+}$ by the relative ionic radii of Pr$^{3+}$ and Ho$^{3+}$. The refinement was performed in the \emph{Spectre} program by a least-squares fitting algorithm in which the experimental uncertainties in the energies and relative intensities of the peaks were used as reciprocal weights. This procedure converged to an excellent fit with $\chi^2 = 3.1$, where $\chi^2$ is the standard normalised goodness-of-fit parameter. Subsequently, a pattern search algorithm \cite{patternsearch} with random starting coefficients was used to explore the $3\times 5! = 360$ remaining possible orderings of the crystal field levels that would also have a ground-state doublet, employing a simplified model with $LS$-coupling and $J$-mixing of only the lowest two $J$-multiplets ($^3H_4$ and $^3H_5$). These solutions were then used as starting-points for least-squares fits to the experimental data performed by {\it Spectre}, but no other acceptable solutions were found.

\begin{table}
\caption{\label{table1} Observed and calculated crystal-field transition energies and intensities of Pr$_2$Sn$_2$O$_7$. Intensities are calculated in the dipole approximation. The best-fit crystal-field parameters used for the calculations are: $B^2_0 = 57.9$, $B^4_0 = 432.8$, $B^4_3 = 161.0$, $B^6_0 = 144.5$, $B^6_3 = -107.5$, $B^6_6 = 192.3$\,meV. Numbers in parentheses indicate errors.}
\begin{ruledtabular}
\begin{tabular}{ccccc}
\textrm{Level}&
\textrm{$E_{\rm obs}$}&
\textrm{$E_{\rm calc}$}&
\textrm{$I_{\rm obs}$}\footnote{Intensity relative to largest inelastic peak.}&
\textrm{$I_{\rm calc}$}\\
\colrule
$\Gamma_3^+$ & 0.0 & 0.0 & $-$ & 3.91\\
$\Gamma_1^+$ & 17.8(4) & 18.0 & 1.00 & 1.00\\
$\Gamma_3^+$  & 57.8(4) & 57.5  & 0.23(6) & 0.34\\
$\Gamma_1^+$  & 82.2(4) & 82.3 & 0.66(7) & 0.62\\
$\Gamma_3^+$  & 100.0(5) & 100.2 & 0.68(7) & 0.65\\
$\Gamma_2^+$  & 115.0(5) & 114.9 & 0.56(7) & 0.41\\
\end{tabular}
\end{ruledtabular}
\end{table}

Table~\ref{table1} gives the best-fit CFPs together with the calculated energies and relative intensities (calculated in the dipole approximation). The calculated values are seen to agree well with the observations. Moreover, the absolute intensities are also in good agreement with the predictions of the model. For example, the sum of the absolute intensities of the five measured inelastic peaks is 425(20) \,mb sr$^{-1}$ f.u.$^{-1}$, which after correction for neutron absorption and self-shielding becomes approximately 530(25) \,mb sr$^{-1}$ f.u.$^{-1}$, and the corresponding calculated sum is 501\,mb sr$^{-1}$ f.u.$^{-1}$. The final CFPs are in close agreement with the scaled parameters determined for Ho$_{2}$Ti$_{2}$O$_{7}$ in Ref. \onlinecite{HoTOCEF}. Based on our extensive search procedure and the closeness of the best-fit crystal field to that in Ho$_{2}$Ti$_{2}$O$_{7}$ we are confident that we have found a unique solution for the crystal field Hamiltonian of Pr$_{2}$Sn$_{2}$O$_{7}$.

\begin{figure}
\includegraphics[width=0.5\textwidth]{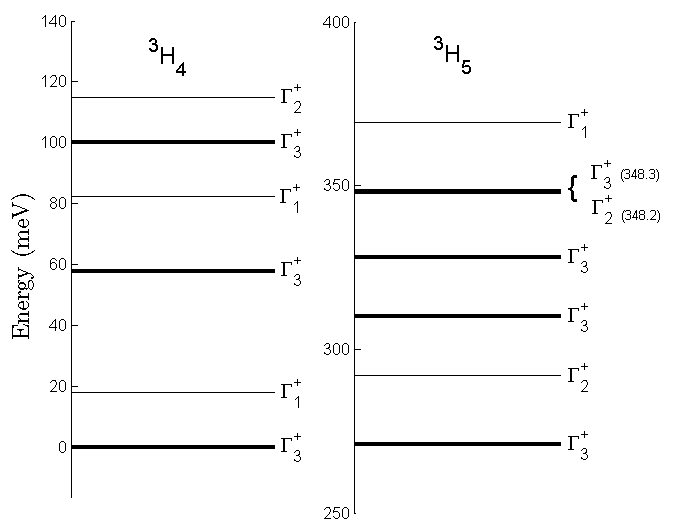}
\caption{\label{cflevels} CF energy level scheme for ground and first excited multiplets of the Pr$^{3+}$ ion in Pr$_{2}$Sn$_{2}$O$_{7}$. Plain and bold lines indicate singlets and doublets, respectively. }
\end{figure}

The energy level scheme for the ground state J-multiplet is shown on the left in Fig.~\ref{cflevels}. The largest eigenvector components of the ground and first excited levels levels, written in terms of the $|^{2S+1}L_J,M_J\rangle$ basis, are found to be
\begin{widetext}
\begin{equation}
\begin{split}
\Gamma_3^+ (0\,{\rm meV})   = \ &  0.88|^3H_4,\pm 4 \rangle \pm  0.41|^3H_4,\pm 1\rangle - 0.14|^3H_4,\mp 2\rangle \nonumber\\
 & + 0.15|^1G_4, \pm 4\rangle \pm 0.07|^1G_4, \pm 1\rangle \mp 0.07|^3H_5,\pm 4\rangle \pm 0.06|^3H_5,\mp 2\rangle     \nonumber\\[5pt]
 \Gamma_1^+ (18\,{\rm meV})  =  \ &  0.14 |^3H_4,3 \rangle + 0.96|^3H_4,0 \rangle - 0.14|^3H_4,-3 \rangle + 0.17|^1G_4,0 \rangle + 0.08|^3F_2,0 \rangle  \nonumber
\end{split}
\label{eq:wavefunctions}
\end{equation}
\end{widetext}

\begin{figure}
\includegraphics[width=0.45\textwidth]{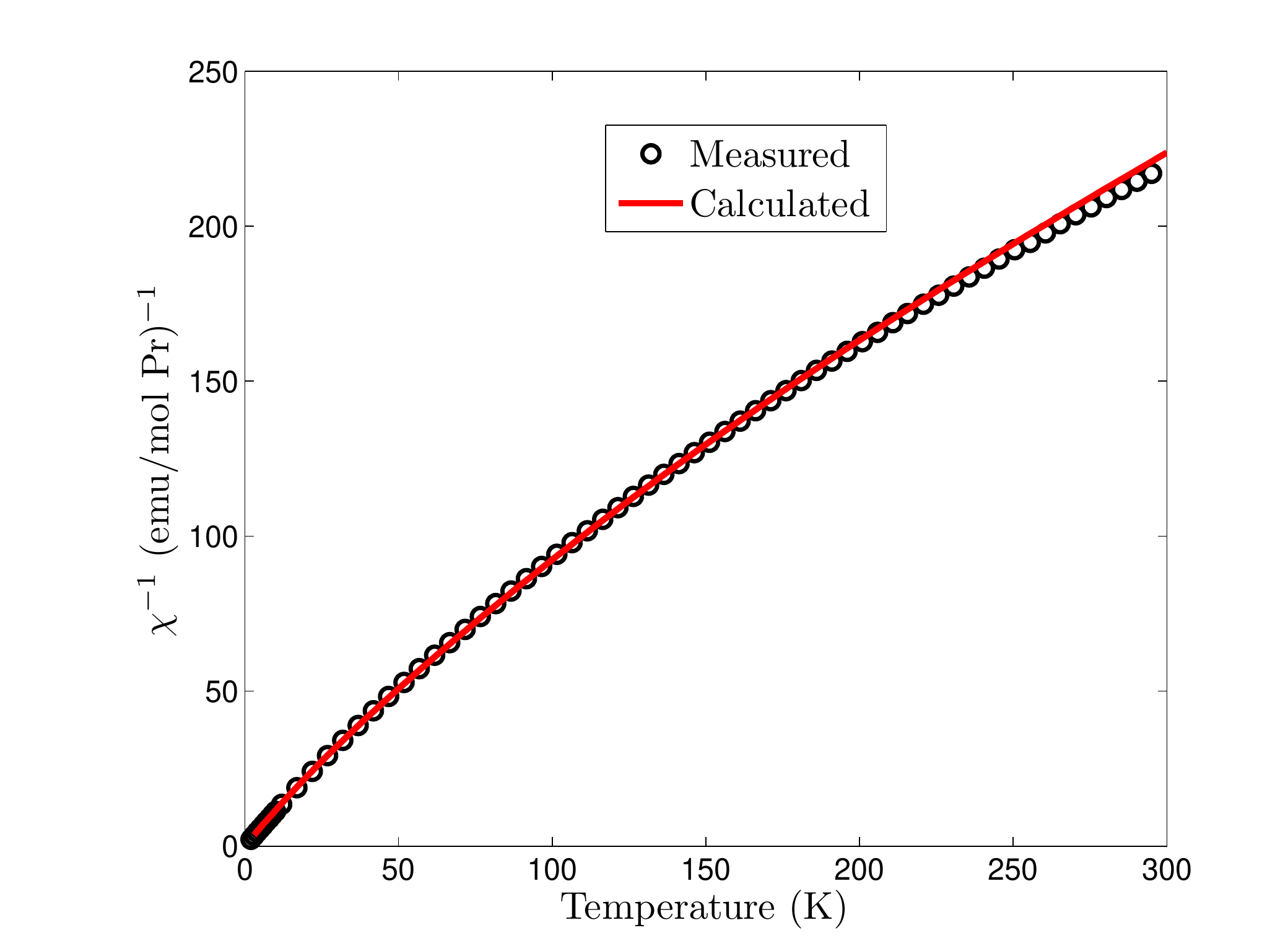}
\caption{\label{susceptibility} Powder-averaged magnetic susceptibility of Pr$_{2}$Sn$_{2}$O$_{7}$. The circles indicate our measurements, and the solid line is calculated from the crystal field model. }
\end{figure}

In Fig.~\ref{intermultiplet} we have plotted the cross-section for the $^3H_4$ $\rightarrow$ $^3H_5$ intermultiplet transitions calculated from our model, and the structure of the excited multiplet is given in Fig.~\ref{cflevels}.  Both the lineshape and absolute intensity match the data very well, providing further support for the crystal-field model. Our model also very accurately reproduces the measured powder-averaged magnetic susceptibility of Pr$_{2}$Sn$_{2}$O$_{7}$, displayed in Fig.~\ref{susceptibility}, which is thus found to be dominated by single-ion physics except at the very lowest temperatures.

\section{Discussion}

Our analysis confirms that the ground state of Pr$^{3+}$ in Pr$_{2}$Sn$_{2}$O$_{7}$ is a non-Kramers doublet of $\Gamma_3^+$ symmetry and that the first excited state is a $\Gamma_1^+$ singlet at 17.8\,meV, as deduced previously by Zhou {\it et al.} (Ref. \onlinecite{ZhouPSO}). However, a peak at 38\,meV that was previously ascribed to a CF transition is identified here as a phonon because it has a strong intensity at large $Q$. According to our analysis, the second excited state is a $\Gamma_3^+$ doublet at 57.8\,meV.

The calculated anisotropy in the susceptibility of Pr$_{2}$Sn$_{2}$O$_{7}$ is $\chi_{\parallel}/\chi_{\perp} \sim 60$ at $T = 10$\,K, where $\chi_{\parallel}$ and $\chi_{\perp}$ are the susceptibilities parallel and perpendicular to the $[111]$ quantisation axis of the crystal field. In the Ho and Dy titanates the corresponding values at 10\,K calculated from a single-ion crystal field model are 350 and 300, respectively \cite{HoTOCEF}. Pr$_{2}$Sn$_{2}$O$_{7}$ is thus highly Ising-like, albeit less than the titanate spin-ice compounds. It should be pointed out that the transverse part of the susceptibility is purely Van-Vleck type, and so the ratio  $\chi_{\parallel} / \chi_{\perp}$ tends to infinity as $T \rightarrow 0$. Since the dominant exchange in Pr$_{2}$Sn$_{2}$O$_{7}$ is ferromagnetic and it exhibits no magnetic order down to 200\,mK, it satisfies the requirements of a spin ice.

We have also calculated from our model the $g$-factor of Pr$_{2}$Sn$_{2}$O$_{7}$ for fields parallel to the quantization axis, which is needed for an effective $S=\frac{1}{2}$ Hamiltonian. We find $g_{\parallel} = 5.17$, with a calculated field-induced moment at low $T$ of 2.6\,$\mu_{\rm B}$ in excellent agreement with measurements \cite{PSOmeasurements}.

 From the coefficients in the ground-state doublet given above, Pr$_{2}$Sn$_{2}$O$_{7}$ is strongly anisotropic, with the normalised expectation value for the z-component (i.e. along the $\langle 111 \rangle$ direction) of the dipole operator $\langle J_z \rangle /J \approx 0.8$ whilst $\langle J_x \rangle /J =\langle J_y \rangle /J =0$ . For comparison, the dipole operator in the Ho and Dy titanates can be extrapolated from the CFPs given by Rosenkrantz et.al (Ref. \onlinecite{HoTOCEF}) to be approximately  0.98 and 0.97, respectively whilst $\langle J_x \rangle /J $ and $\langle J_y \rangle /J $ remain zero. The fact that $\langle J_z \rangle /J$ is substantially less than 1 in  Pr$_{2}$Sn$_{2}$O$_{7}$ indicates that there are significant fluctuations of the moment away from the quantisation axis. Owing to the fact that Pr$^{3+}$ is a non-kramers ion, such fluctuations cannot arise as the result of a linear coupling and must be given by a bilinear operator in the form of e.g. some component of the quadrupole, $\{ J_z,J_\pm \}$. In this case we find that the nonzero component of the quadrupole operator has an expectation value within the ground state  approximately 1/3 the size of $\langle J_z \rangle $.

Finally, it is interesting to quantify the approximation that one would make in neglecting both intermediate coupling and $J$-mixing effects. To this end, we repeated the data analysis using the Stevens operator equivalents $O_q^k$ to describe the crystal field \cite{hutchingsoperatortechniques}, i.e. by diagonalizing $\mathcal{H}_{\rm CF}$ within the pure $LS$-coupling $J=4$ multiplet and neglecting $J$-mixing effects. Weobtain the best-fit parameters$B_0^2 = -733$, $B_0^4 = -36.5 $, $B_3^4 = -383 $, $B_0^6 =  0.278$, $B_3^6 = 0.0328 $, and $B_6^6 = -4.59 $ in units of $\mu$eV which yield the following wave functions of the ground and first excited states
\begin{equation}
\begin{split}
 \Gamma_3^+  =  &  0.93| \pm 4\rangle \pm 0.37|\pm 1 \rangle + 0.05| \mp 2 \rangle  \nonumber \\
 \Gamma_1^+   = &  0.22|3 \rangle + 0.95| 0 \rangle - 0.22| -3 \rangle   \nonumber
\end{split}
\end{equation}
This approach reduces the expectation value of the quadrupole operator by a factor of 3 compared to a full calculation done in intermediate coupling. It is clear, therefore, that the use of intermediate coupling basis states and inclusion of $J$-mixing is important to any theoretical description of quantum effects in this material.

\section{Conclusion}

By fitting inelastic neutron scattering data we have determined that the ground-state of  Pr$^{3+}$ in Pr$_{2}$Sn$_{2}$O$_{7}$ is a well-isolated doublet and extracted the ground and excited-state wave-functions. Additionally, we have shown quantitatively that there are strong transverse fluctuations in the ground state, compared with classical spin ices such as the Dy and Ho titanates. This confirms the prediction that Pr$_2$Sn$_2$O$_7$ is a very promising candidate for quantum spin ice.

\emph{Note added}: subsequent to the submission of this article, we were made aware of a new publication by Kimura et. al. (Ref. \onlinecite{Pr2Zn2O7seminal}) in which the authors successfully determine the crystal field of the isostructural compound Pr$_2$Zn$_2$O$_7$, using the Stevens operator formalism. The results of  Kimura et. al. are qualitatively similar to those determined in this paper for Pr$_{2}$Sn$_{2}$O$_{7}$ when also using Stevens operator approach.

\begin{acknowledgments}
This work was financially supported by the U.K. Engineering and Physical Sciences Research Council under Grant No. EP/H014934/1. We would like to thank Michel Gingras for helpful suggestions that aided the construction of this paper.
\end{acknowledgments}

\bibliography{masterpaperlist}

\end{document}